\documentstyle[11pt,epsf]{article}
\input epsf
\title{Stationary Axisymmetric Configuration of the Resistive Thick
Accretion Tori around a Schwarzschild Black Hole}

\author
       {M. Shaghaghian \\
        Department of Physics, Science and Research Branch, Islamic Azad University, Fars, Iran\\
        Department of Physics, Shiraz Branch, Islamic Azad University, Shiraz, Iran\\
        }

  \hoffset =
-2cm \textwidth = 170mm

\begin{document}

\maketitle

\abstract We examine a thick accretion disc in the presence of
external gravity and intrinsic dipolar magnetic field due to a
non-rotating central object. In this paper, we generalize the
Newtonian theory of stationary axisymmetric resistive tori of
Tripathy, Prasanna $\&$ Das (1990) by including the fully general
relativistic features. If we are to obtain the steady state
configuration, we have to take into account the finite resistivity
for the magnetofluid in order to avoid the piling up of the field
lines anywhere in the accretion discs. The efficient value of
conductivity must be much smaller than the classical conductivity to
be astrophysically interesting. The accreting plasma in the presence
of an external dipole magnetic field gives rise to a current in the
azimuthal direction. The azimuthal current produced due to the
motion of the magnetofluid modifies the magnetic field structure
inside the disc and generates a poloidal magnetic field for the
disc. The solutions we have found show that the radial inflow,
pressure and density distributions are strongly modified by the
electrical conductivity both in relativistic and Newtonian regimes.
However, the range of conductivity coefficient is different for both
regimes, as well as that of the angular momentum parameter and the
radius of the innermost stable circular orbit. Furthermore, it is
shown that the azimuthal velocity of the disc which is not dependent
on conductivity is sub-Keplerian in all radial distances for both
regimes. Owing to the presence of pressure gradient and magnetic
forces. This work may also be important for the general relativistic
computational magnetohydrodynamics that suffers from the lack of
exact analytic solutions that are needed to test computer codes.
\section{Introduction}

After the discovery of quasars in 1963, black holes have attracted a
lot of attention in high-energy astrophysics as energizers. Release
of gravitational energy through the accretion of matter on to a
massive black hole is widely regarded as the most plausible origin
of the enormous luminosities of quasars. Spherical accretion on to
massive black holes is inefficient in releasing this energy, and any
accreting matter is likely to have the relative high angular
momentum content. Hence, it is common to invoke the presence of an
accretion disc, that could be thin or thick depending upon their
geometrical shapes.

The standard theory of thin accretion discs is mostly based on the
fundamental paper of Shakura $\&$ Sunyaev (1973, hereafter SS73).
Because of its relative simplicity and successful applicability,
this theory has achieved the status of a textbook paradigm. It
introduces a Keplerian rotating gas flow which is stationary,
rotationally symmetric, geometrically thin and optically thick. This
flow has subcritical accretion rate with vertical hydrostatic
equilibrium, negligible pressure gradients and insignificant infall
velocity. Moreover, it makes the physically reasonable and
enormously productive Ansatz about the turbulent viscosity which is
parametrized with an alpha parameter. The SS73 model has been found
to be quite successful in reproducing at least the gross features of
the observations (see e.g. Frank, King $\&$ Raine 2002).

Novikov $\&$ Thorne (1973) then included general relativity and
extended the SS73 model to relativistic flows around the rotating
black holes. Studies wishing to model more accurately phenomena
beyond the SS73 model like time-dependent behavior, radiatively
driven mass loss, various instabilities etc. have stretched the SS73
model over the range of its limits. For a review of such
applications see e.g. Lin $\&$ Papaloizou (1996).

The common practice to treat accretion discs using the SS73 paradigm
has always been accompanied by vertical averaging or in other word
integrating the flow equations in the vertical direction (i.e.
parallel to the rotation axis). Vertical averaging is a standard
approximation in accretion disc theory since a long time ago, owing
to the simplifications it introduces (Mineshige $\&$ Umemura 1997;
Gammie $\&$ Popham 1998; Tsuribe 1999). The physical motivation
behind this approximation is that the vertical thickness of the disc
is usually much smaller than the local radius, so that the flow
velocities are likely to be more or less independent of height. It
is then reasonable to expect that very little is lost by integrating
out the vertical coordinate and employing the vertical hydrostatic
equilibrium.

In this paper we avoid the height-integration approximation and
subsequently the vertical hydrostatic equilibrium. Instead, we set
up the exact flow equations for steady axisymmetric flow in the $r
\theta$ plane. Consequently, the resulting vertical pressure
gradient may cause the disc to bulge and grow outwards from the
equatorial plane, becoming a thick disc in which the vertical
dimension of the disc is comparable with its radial size. For a
geometrically thick accretion disc that the pressure forces play an
essential role in its equilibrium structure, pressure provides a
substantial support in the radial as well as vertical direction. As
a result, gravity of the central star is no longer balanced by
centrifugal force. Thus, the rotation law may be far from Keplerian
distribution here.

The basic equations governing the motion of an axisymmetric
stationary magnetofluid disc around a compact object in a curved
space-time background are given by Prasanna, Tripathy $\&$ Das
(1989, hereafter PTD89). The behaviour of a fluid in the presence of
electromagnetic fields is governed to a large extent by the
magnitude of the electrical conductivity. In most of the
astrophysical phenomena, the ideal MHD approximation, in which the
conductivity is actually assumed to be infinite, represents a very
good approximation. In this case, the magnetic flux is conserved and
the magnetic field is frozen in the fluid, being simply advected
with it. Ideal MHD equations neglect any effect of resistivity on
the dynamics. However, in cold, dense plasmas such as might be
expected at the centres of protostellar discs (Stone et al. 2000;
Fleming $\&$ Stone 2003), discs in dwarf nova systems (Gammie $\&$
Menou 1998), the ionization fraction may become so small that this
approximation no longer holds and the conductivity must be assumed
to be finite. Moreover, in practice, even in the scenarios of hot
plasmas like accretion discs around black holes (Kudoh $\&$ Kaburaki
1996), and in Galactic centre (Melia $\&$ Kowalenkov 2001; Kaburaki
et al. 2010), there will be spatial regions where the electrical
conductivity is finite too. In these instances, the resistive
effects, most notably, magnetic reconnection will be expected to
occur in reality. It will provide an important contribution to the
energy losses from the system. Inclusion of a finite resistivity is
particularly essential for a non-viscous disc to liberate
gravitational energy.

The first analytical equilibrium solution including the conductivity
for the plasma was obtained by Kaburaki (1986, 1987). Although his
equilibrium solutions were not self-consistent, but he showed that
similar to the standard viscous disc model of SS73, about one half
of the gravitational energy is released in the magnetized disc
through the Joule dissipation. As a consequence, the magnetic stress
can take the place of viscous stress in the standard disc model, and
extracts angular momentum from the disc.

In recent decades, plenty of studies on the dynamics of accretion
discs have been carried out by applying some simplifying assumptions
on the general equations derived by PTD89. The most usual of these
assumptions are thin disc approximation (Prasanna $\&$ Bhaskaran
1989; Prasanna 1989; Bhaskaran $\&$ Prasanna 1990) and flow
restriction only to the azimuthal component, with the inherent
assumption that the radial and meridional components are negligible
in comparison with the azimuthal one (Banerjee et al. 1997).
Furthermore, Newtonian limit of relativistic equations of PTD89 for
a thick resistive plasma disc surrounding a non-rotating black hole
with an intrinsic dipolar magnetic field has been examined by
Tripathy, Prasanna $\&$ Das (1990, hereafter TPD90). Despite the
fact that for the first time, they took into consideration all three
components of the flow velocity of matter, but yet another important
aspect in thick accretion disc theory around the compact objects is
left. Lacuna is the generalization of the Newtonian analysis of
TPD90 to general relativistic formalism wherein the space-time
curvature produced by the strong gravitational field of the central
body introduces the new features.

In this paper we fill in this lacuna and follow the model drawn by
TPD90, but put aside the Newtonian limit and work in the fully
relativistic framework. We investigate the general relativistic
effects by choosing the Schwarzschild geometry and ignore the
self-gravity contribution of the disc. Further, we consider the
central object to possess a poloidal magnetic field which has the
usual dipolar form in the asymptotic limit of the Schwarzschild
metric. We are not interested in outflowing motion from the surface
of the disc. Consequently, neglecting the meridional flow (i.e.
$V^{\theta}=0$), seems an admissible approximation as a simplifying
assumption in solving equations (Gu et al. 2009). Therefore, we
proceed to study a pressure-supported, accreting, stationary,
axisymmetric, conducting, magnetized thick disc around a compact
object within the relativistic domain without shear viscosity and
meridional flow.

The objective of this work is to attempt to understand the effect of
electrical conductivity and the other free parameters on the
dynamics of disc with approximations that allow the problem to be
treated mainly by analytical methods as far as possible. A brief
outline of this paper is as follows: In Section 2, we derive a set
of dynamical basic equations in Schwarzschild background and discuss
the magnetic field configuration. In Section 3, the analytical and
numerical solutions of the relativistic disc are obtained and the
effects of various parameters involved in the governing equations on
these solutions are investigated. Newtonian limit of the
relativistic solutions and comparison of these two sets of solutions
are practiced in Section 4. Section 5 summarizes our conclusions.

\section{GENERAL FORMULATION}
\subsection{Basic Equations}

We are interested in relativistic magnetized flow accreted from the
source of plasma around a non-rotating black hole in the form of a
thick disc. Our desired magnetofluid disc is in stationary
($\partial_t\equiv 0$) and axisymmetric ($\partial_{\varphi}\equiv
0$). Further, it is not massive in comparison with the central
compact object. Hence, the self-gravity of the disc is considered to
be negligible and the space-time structure supporting the disc is
determined entirely by the central body.  Besides, energy of the
electromagnetic field is regarded to be negligible as compared to
the energy associated with the mass of the central star. It means
the electromagnetic fields do not influence the geometry, but they
can be modified by the background geometry that is defined by the
Schwarzschild metric
\begin{eqnarray*}
ds^2=\left(1-\frac{2m}{r}\right)c^2
dt^2-\left(1-\frac{2m}{r}\right)^{-1} dr^2
-r^2\left(d\theta^2+\sin^2\theta d\varphi^2\right).
\end{eqnarray*}
The fundamental unit of length is $m=\frac{G M}{c^2}$, with $G$ as
the universal gravitational constant, $M$ mass of the central
object, and $c$ the speed of light. The Schwarzschild radial
coordinate $r$ may be normalized with respect to $m$.

In general, motion of the magnetized plasma is described by three
sets of general relativistic MHD equations. These are the
energy-momentum conservation laws
\begin{equation}
\label{Conserv of T} T^{ij}_{\ ;j}=0.
\end{equation}
and Maxwell equations
\begin{eqnarray}
\label{Max Equ1} &&F^{ij}_{\ ;j}=-\frac{4\pi}{c} J^i,\\[.3cm]
\label{Max Equ2} &&F_{ij,k}+F_{ki,j}+F_{jk,i}=0,
\end{eqnarray}
along with the generalized Ohm law
\begin{eqnarray}\label{Ohm law}
J^i=\sigma F^i_{\ k} u^k.
\end{eqnarray}
It is worth noting here that semicolon denotes a covariant
derivative with respect to $x^j$. For a fluid endowed with a
magnetic field, stress-energy tensor $T^{ij}$ is obtained by adding
the energy-momentum tensor of the fluid
\begin{eqnarray*}
T^{ij}_{Fluid}=\left(\rho+\frac{p}{c^2}\right) u^i
u^j-\frac{p}{c^2}\ g^{ij},
\end{eqnarray*}
to that of the electromagnetic field
\begin{eqnarray*}
T^{ij}_{Em}=-\frac{1}{4\pi c^2}\left(F^{ik} \ F^j_{\ k}-\frac{1}{4}\
g^{ij}\ F_{kl}\ F^{kl}\right),
\end{eqnarray*}
as
\begin{eqnarray*}
T^{ij}=T^{ij}_{Fluid}+T^{ij}_{Em}.
\end{eqnarray*}
It consists of a perfect fluid with the rest-mass density $\rho$,
the pressure $p$, the four-velocity $u^i$ and an electromagnetic
field tensor $F^{ij}$ satisfying Maxwell equations. The other fluid
variables are the electric four-current density $J$ and the electric
conductivity $\sigma$ which is assumed constant for simplicity. We
express the dynamical equations in terms of physical quantities by
writing them in the orthonormal tetrad frame appropriate to the
Schwarzschild metric
\begin{eqnarray*}
\lambda^{i}_{(\alpha)}=diag\left[\left(1-\frac{2m}{r}\right)^{-1/2},\left(1-\frac{2m}{r}\right)^{1/2},\frac{1}{r},\frac{1}{r
\sin \theta}\right],
\end{eqnarray*}
satisfying
$$\lambda^i_{(a)}\ \lambda^j_{(b)}\ g_{ij}=\eta_{(a)(b)},$$
where $g_{ij}$ and $\eta_{(a)(b)}$ are the metric and Minkowski
tensors, respectively. All global variables are then defined in
local Lorentz frame as follows:
\begin{eqnarray*}
F_{(\alpha)(\beta)}&=&\lambda^i_{\ (\alpha)}\ \lambda^j_{\ (\beta)} F_{ij},\\
J^{(\alpha)}&=&\lambda^{(\alpha)}_{\ \ i} J^i,\\
E_{(\alpha)}&=&F_{(\alpha)(t)},\\
B_{(\alpha)}&=&\epsilon_{\alpha\beta\gamma}F_{(\beta)(\gamma)},
\end{eqnarray*}
where $\epsilon_{\alpha\beta\gamma}$ is the Levi-Civita symbol.
Using these definitions, one can express the spatial 3-velocity
$V^{\alpha}$ defined through the relation
$u^{\alpha}=\frac{u^0}{c}V^{\alpha}$, in terms of local Lorentz
components as given by
\begin{eqnarray*}
V^{(r)}&=&\left(1-\frac{2m}{r}\right)^{-1} V^r,\\
V^{(\theta)}&=&r\left(1-\frac{2m}{r}\right)^{-1/2} V^{\theta},\\
V^{(\varphi)}&=&r\sin\theta\left(1-\frac{2m}{r}\right)^{-1/2}V^{\varphi}.
\end{eqnarray*}

There exists a minimal radius at which stable circular motion is
still possible for the plasma orbiting the central black hole. It
defines the so-called innermost stable circular orbit (ISCO) in
given background. For the Schwarzschild geometry, the radius of ISCO
equals to $6 m$ (Jefremov, Tsupko $\&$ Bisnovatyi 2015). Although,
if the surface magnetic field of the central black hole is not too
high, it can reach well within the usual $6m$ limit, almost upto
$3m$ (Bhaskaran $\&$ Prasanna 1990). As a result, we presume the
disc spreads in radial direction from $6m$ to $50m$ and in
meridional direction has the angular thickness $\pi/3$ on either
side of the equator.

To determine the configuration of magnetic field lines and the
velocity of MHD flows streaming along each magnetic field line, we
must solve the basic equations (\ref{Conserv of T}) - (\ref{Ohm
law}) self-consistently. We expand the equations in local Lorentz
frame noting to the fact that the Roman indices run from 0 to 3 and
the Greek ones run from 1 to 3. Besides, we adopt the standard
convention for the summation over repeated indices. The zeroth
component of equation (\ref{Conserv of T}) is the continuity
equation
\begin{eqnarray}\label{Continuity equ1}
&&\left(\rho+\frac{p}{c^2}\right) \left\{\frac{\partial
V^{(r)}}{\partial
r}+\frac{2}{r}V^{(r)}+\frac{1}{r}\left(1-\frac{2m}{r}\right)^{-1/2}
\left[\frac{\partial V^{(\theta)}}{\partial \theta}+\cot\theta V^{(\theta)}\right]\right\}\nonumber\\[.1cm]
&&+V^{(r)}\frac{\partial}{\partial r}\left(\rho-\frac{p}{c^2}\right)+\left(1-\frac{2m}{r}\right)^{-1/2}\frac{V^{(\theta)}}{r}\frac{\partial}{\partial \theta}\left(\rho-\frac{p}{c^2}\right)\nonumber\\[.1cm]
&&+\left(1-\frac{2m}{r}\right)^{-1/2}\frac{2}{c^3}\left[B_{(\theta)}V^{(r)}J^{(\varphi)}-E_{(r)}V^{(r)}J^{(t)}
-B_{(r)} V^{(\theta)}J^{(\varphi)}-E_{(\theta)} V^{(\theta)}
J^{(t)}\right]=0,
\end{eqnarray}
and its spatial components provide the momentum equations
\begin{eqnarray}\label{Radial equ1}
&&\frac{\left(\rho+\frac{p}{c^2}\right)}{\left(1-\frac{V^2}{c^2}\right)}\left[V^{(r)}\frac{\partial V^{(r)}}{\partial r}+\left(1-\frac{2m}{r}\right)^{-1/2}\frac{V^{(\theta)}}{r}\frac{\partial V^{(r)}}{\partial \theta}\right.\nonumber\\[.1cm]
&&\left.+\frac{m c^2}{r^2}\left(1-\frac{2m}{r}\right)^{-1}\left\{1-\frac{\left[V^{(r)}\right]^2}{c^2}\right\}\right.\nonumber\\[.1cm]
&&\left.-\frac{1}{r}\left\{[V^{(\theta)}]^2+[V^{(\varphi)}]^2\right\}\right]+\frac{\partial p}{\partial r}\nonumber\\[.1cm]
&&+\left(1-\frac{2m}{r}\right)^{-1/2}\frac{1}{c}\left[E_{(r)}J^{(t)}-B_{(\theta)}J^{(\varphi)}\right]=0,
\end{eqnarray}

\begin{eqnarray}\label{Meridional equ1}
&&\frac{\left(\rho+\frac{p}{c^2}\right)}{\left(1-\frac{V^2}{c^2}\right)}\left\{V^{(r)}\frac{\partial V^{(\theta)}}{\partial r}+\left(1-\frac{2m}{r}\right)^{-1/2}\frac{V^{(\theta)}}{r}\frac{\partial V^{(\theta)}}{\partial \theta}\right.\nonumber\\[.1cm]
&&\left.+\frac{\left(1-\frac{3m}{r}\right)}{\left(1-\frac{2m}{r}\right)}\frac{V^{(r)}V^{(\theta)}}{r}-\cot\theta\left(1-\frac{2m}{r}\right)^{-1/2}\frac{\left[V^{(\varphi)}\right]^2}{r}\right\}\nonumber\\[.1cm]
&&+\left(1-\frac{2m}{r}\right)^{-1/2}\frac{1}{r}\frac{\partial p}{\partial\theta}\nonumber\\[.1cm]
&&+\left(1-\frac{2m}{r}\right)^{-1/2}\frac{1}{c}\left[E_{(\theta)}J^{(t)}+B_{(r)}J^{(\varphi)}\right]=0,
\end{eqnarray}

\begin{eqnarray}\label{Azimuthal equ1}
&&V^{(r)}\frac{\partial V^{(\varphi)}}{\partial r}+\left(1-\frac{2m}{r}\right)^{-1/2}\frac{V^{(\theta)}}{r}\frac{\partial V^{(\varphi)}}{\partial \theta}\nonumber\\[.1cm]
&&+\frac{\left(1-\frac{3m}{r}\right)}{\left(1-\frac{2m}{r}\right)}\frac{V^{(r)}V^{(\varphi)}}{r}+\left(1-\frac{2m}{r}\right)^{-1/2}\cot\theta\frac{V^{(\theta)}V^{(\varphi)}}{r}=0.\qquad
\end{eqnarray}
The Maxwell equations (\ref{Max Equ1}) and (\ref{Max Equ2}) can be
expanded as
\begin{eqnarray}\label{Maxwell equ1}
\frac{\partial}{\partial\theta}\left(\sin\theta\
B_{(\varphi)}\right)=-\frac{4\pi}{c} \, r \sin\theta J^{(r)},
\end{eqnarray}
\begin{eqnarray}\label{Maxwell equ2}
\frac{\partial}{\partial r}\left[r \left(1-\frac{2m}{r}\right)^{1/2}
B_{(\varphi)}\right]=\frac{4\pi}{c} \, r J^{(\theta)},
\end{eqnarray}
\begin{eqnarray}\label{Maxwell equ3}
\frac{\partial}{\partial r}\left[r \left(1-\frac{2m}{r}\right)^{1/2}
B_{(\theta)}\right]-\frac{\partial
B_{(r)}}{\partial\theta}=-\frac{4\pi}{c} \, r J^{(\varphi)},
\end{eqnarray}
\begin{eqnarray}\label{Maxwell equ4}
\left(1-\frac{2m}{r}\right)^{1/2}\sin\theta\frac{\partial}{\partial
r}[r^2 E_{(r)}]+r\frac{\partial}{\partial \theta}[\sin\theta
E_{(\theta)}]= -\frac{4\pi}{c}\ r^2\sin\theta J^{(t)},
\end{eqnarray}
\begin{eqnarray}\label{Maxwell equ5}
\sin\theta\frac{\partial}{\partial r}\left[r^2 B_{(r)}\right]+r
\left(1-\frac{2m}{r}\right)^{-1/2}\frac{\partial}{\partial\theta}\left[\sin\theta
\ B_{(\theta)}\right]=0,
\end{eqnarray}
\begin{eqnarray}\label{Maxwell equ6}
\frac{\partial}{\partial r}\left[r \left(1-\frac{2m}{r}\right)^{1/2}
E_{(\theta)}\right]-\frac{\partial E_{(r)}}{\partial \theta}=0.
\end{eqnarray}
At the first glance, the set of basic equations (\ref{Continuity
equ1}) -(\ref{Maxwell equ6}) seems rather complex to solve, and in
general requires the use of plausible simplifying assumptions. As a
consequence of axisymmetry of the problem, it looks reasonable to
vanish the toroidal component of the electromagnetic field
($E_{(\varphi)}=B_{(\varphi)}=0$). This, when used in the Maxwell
equations (\ref{Maxwell equ1}) and (\ref{Maxwell equ2}) leads to the
omission of the poloidal current density ($J^{(r)}=J^{(\theta)}=0$).
Afterwards, the Ohm law (equation \ref{Ohm law}) yields
\begin{eqnarray}\label{Er1}
E_{(r)}=B_{(\theta)}\frac{V^{(\varphi)}}{c},
\end{eqnarray}
\begin{eqnarray}\label{Etheta1}
E_{(\theta)}=-B_{(r)}\frac{V^{(\varphi)}}{c},
\end{eqnarray}
\begin{eqnarray}\label{Jf1}
J^{(\varphi)}=-\frac{\sigma}{c} \, u^0
\left(1-\frac{2m}{r}\right)^{1/2}
\left[B_{(\theta)}V^{(r)}-B_{(r)}V^{(\theta)}\right],
\end{eqnarray}
\begin{eqnarray}\label{Jt1}
J^{(t)}=-\frac{\sigma}{c} \, u^0 \left(1-\frac{2m}{r}\right)^{1/2}
\left[E_{(r)}V^{(r)}+E_{(\theta)}V^{(\theta)}\right]=\frac{J^{(\varphi)}
V^{(\varphi)}}{c}.
\end{eqnarray}
Defining the total derivative
\begin{eqnarray*}
d\equiv V^{(r)}\frac{\partial}{\partial
r}+\left(1-\frac{2m}{r}\right)^{-1/2}\frac{V^{(\theta)}}{r}\frac{\partial}{\partial
\theta},
\end{eqnarray*}
we rewrite the equations (\ref{Continuity equ1}) - (\ref{Azimuthal
equ1}) as
\begin{eqnarray}\label{Continuity equ2}
&&\left(\rho+\frac{p}{c^2}\right)\frac{1}{r^2
\sin\theta}\left\{\frac{\partial}{\partial r}\left[r^2 \sin\theta
V^{(r)}\right]+
\right.\nonumber\\[.1cm]
&&\left.\frac{\partial}{\partial \theta}\left[r \sin\theta
\left(1-\frac{2m}{r}\right)^{-1/2} V^{(\theta)}\right]\right\}
+d\left(\rho-\frac{p}{c^2}\right)\nonumber\\[.1cm]
&&-\frac{2}{\sigma c^2
u^0}\left(1-\frac{2m}{r}\right)^{-1}[J^{(\varphi)}]^2\left\{1-\frac{[V^{(\varphi)}]^2}{c^2}\right\}=0,
\end{eqnarray}

\begin{eqnarray}\label{Radial equ2}
&&\left(\rho+\frac{p}{c^2}\right)\left(1-\frac{V^2}{c^2}\right)^{-1}\left[d V^{(r)}+\frac{m c^2}{r^2}\left(1-\frac{2m}{r}\right)^{-1}\times\right.\nonumber\\[.1cm]
&&\left.\left\{1-\frac{[V^{(r)}]^2}{c^2}\right\}-\frac{1}{r}\left\{[V^{(\theta)}]^2+[V^{(\varphi)}]^2\right\}\right]+\frac{\partial p}{\partial r}\nonumber\\[.1cm]
&&-\left(1-\frac{2m}{r}\right)^{-1/2} \frac{1}{c} \, B_{(\theta)}
J^{(\varphi)}\left\{1-\frac{[V^{(\varphi)}]^2}{c^2}\right\}=0,
\end{eqnarray}

\begin{eqnarray}\label{Meridional equ2}
&&\left(\rho+\frac{p}{c^2}\right)\left(1-\frac{V^2}{c^2}\right)^{-1}\left\{d V^{(\theta)}+\frac{\left(1-\frac{3m}{r}\right)}{\left(1-\frac{2m}{r}\right)}\frac{V^{(r)}V^{(\theta)}}{r}\right.\nonumber\\[.1cm]
&&\left.-\cot\theta\left(1-\frac{2m}{r}\right)^{-1/2}\frac{\left[V^{(\varphi)}\right]^2}{r}\right\}+\left(1-\frac{2m}{r}\right)^{-1/2}\frac{1}{r}\frac{\partial p}{\partial\theta}\nonumber\\[.1cm]
&&+\left(1-\frac{2m}{r}\right)^{-1/2}\frac{1}{c} \, B_{(r)}
J^{(\varphi)}\left\{1-\frac{[V^{(\varphi)}]^2}{c^2}\right\}=0,
\end{eqnarray}

\begin{eqnarray}\label{Azimuthal equ2}
&&d V^{(\varphi)}
+\left(1-\frac{3m}{r}\right)\left(1-\frac{2m}{r}\right)^{-1}\frac{V^{(r)}V^{(\varphi)}}{r}\nonumber\\[.1cm]
&&+\cot\theta\left(1-\frac{2m}{r}\right)^{-1/2}\frac{V^{(\theta)}
V^{(\varphi)}}{r}=0.
\end{eqnarray}
Equation (\ref{Azimuthal equ2}) can be rewritten as
\begin{eqnarray}\label{Azimuthal equ3}
&&d\, \left[r \sin\theta \left(1-\frac{2m}{r}\right)^{-1/2}
V^{(\varphi)}\right]=0.
\end{eqnarray}
Integrating equation (\ref{Azimuthal equ3}), the azimuthal velocity
is achieved
\begin{eqnarray}\label{azimuthal velocity}
V^{(\varphi)}=\frac{L}{r\sin\theta}\left(1-\frac{2m}{r}\right)^{1/2},
\end{eqnarray}
where $L$ is a constant of integration and is defined as $L=l c m$,
wherein $l$ is called the angular momentum parameter.

The poloidal component of the disc's magnetic field is given by
equations (\ref{Maxwell equ5}) and (\ref{Maxwell equ6}).
Self-consistent solution of these two equations can be achieved by
combining these equations together as
\begin{eqnarray}\label{Magnetic Configuration}
B_{(r)}\left(1-\frac{2m}{r}\right)^{-1/2}
\left(1-\frac{3m}{r}\right)+B_{(\theta)}\frac{\cos \theta}{\sin
\theta}=0.
\end{eqnarray}
One admissible solution set for the magnetic field that satisfy the
equation (\ref{Magnetic Configuration}) is given by
\begin{eqnarray}\label{Br}
B_{(r)}=-B_1 \,
r^{k-1}\left(1-\frac{2m}{r}\right)^{-\frac{k}{2}}\sin^{k-1}\theta \,
\cos\theta,
\end{eqnarray}
\begin{eqnarray}\label{Btheta}
B_{(\theta)}=B_1 \,
r^{k-1}\left(1-\frac{3m}{r}\right)\left(1-\frac{2m}{r}\right)^{-\frac{k+1}{2}}\sin^{k}\theta,
\end{eqnarray}
wherein $k$ and $B_1$ are constant. Relations (\ref{Er1}) and
(\ref{Etheta1}) give the poloidal components of the electric field
too
\begin{eqnarray}\label{Er2}
E_{(r)}=\frac{L}{c} B_1 \,
r^{k-2}\left(1-\frac{3m}{r}\right)\left(1-\frac{2m}{r}\right)^{-\frac{k}{2}}\sin^{k-1}\theta,
\end{eqnarray}
\begin{eqnarray}\label{Etheta2}
E_{(\theta)}=\frac{L}{c} B_1 \,
r^{k-2}\left(1-\frac{2m}{r}\right)^{-\frac{k-1}{2}}\sin^{k-2}\theta
\, \cos\theta.
\end{eqnarray}
Current density components can be achieved by the other unused
Maxwell equations (\ref{Maxwell equ3}) and (\ref{Maxwell equ4})
\begin{eqnarray}\label{Jf2}
&&J^{(\varphi)}=\frac{c}{4\pi} B_1 \, r^{k-2}\sin^{k}\theta \left(1-\frac{2m}{r}\right)^{-\frac{k}{2}}\times\nonumber\\[.1cm]
&&\left[\left(1-\frac{3m}{r}\right) -k
\frac{\left(1-\frac{3m}{r}\right)^2}{\left(1-\frac{2m}{r}\right)}+(1-k)\cot^2\theta\right],
\end{eqnarray}
\begin{eqnarray}\label{Jt2}
&&J^{(t)}=\frac{L}{4\pi} B_1 \, r^{k-3}\sin^{k-1}\theta \left(1-\frac{2m}{r}\right)^{-\frac{k-1}{2}}\times\nonumber\\[.1cm]
&&\left[\left(1-\frac{3m}{r}\right) -k
\frac{\left(1-\frac{3m}{r}\right)^2}{\left(1-\frac{2m}{r}\right)}+(1-k)\cot^2\theta\right].
\end{eqnarray}
As seen, two different definitions have been obtained for the
current density $J^{(\varphi)}$ (equations \ref{Jf1} and \ref{Jf2}).
Evidently, they have to be consistent
\begin{eqnarray}\label{Consistency relation}
\left(1-\frac{3m}{r}\right) V^{(r)}+\left(1-\frac{2m}{r}\right)^{1/2} \cot\theta \, V^{(\theta)}=\frac{c^2}{4\pi\sigma u^0}\frac{1}{r}\times\nonumber\\[.2cm]
\left[k
\frac{\left(1-\frac{3m}{r}\right)^2}{\left(1-\frac{2m}{r}\right)}-\left(1-\frac{3m}{r}\right)
+(k-1)\cot^2\theta\right].
\end{eqnarray}
In deriving this consistency equation, the components of the
poloidal magnetic fields, equations (\ref{Br}) and (\ref{Btheta}),
have been substituted. To sum up, the remaining basic equations that
have not been solved yet, are listed as equations (\ref{Continuity
equ2}) - (\ref{Meridional equ2}) and (\ref{Consistency relation})
including the undetermined physical variables (i.e.
$V^{(r)}$,$V^{(\theta)}$, $\rho$ and $p$).

\subsection{Magnetic Field Configuration}
As a matter of fact, the magnetic field in the MHD equations of the
surrounding space of the central black hole consists generally of
two parts:
$$\textbf{B}=\textbf{B}^S+\textbf{B}^D.$$
$\textbf{B}^S$ represents the external magnetic field generated by
the current streaming outside the event horizon of the central black
hole, while $\textbf{B}^D$ is the disc field caused by the current
flowing in the accretion disc. Roughly speaking, the region where
$\left|\textbf{B}^S\right|\geq\left|\textbf{B}^D\right|$ may be
called the magnetosphere. Magnetodisc is also the region where
$\left|\textbf{B}^S\right|<<\left|\textbf{B}^D\right|$. Its
structure is maintained by the Lorentz force acting on the current
there. Therefore, within the disc, $\textbf{B}$ can be replaced by
$\textbf{B}^D$ with sufficient accuracy. Indeed, the magnetic field
appearing in the equations of the previous sections is just the same
as $\textbf{B}^D$, that its superscript $D$ has been dropped for
simplicity. From now on, we use superscript only for the external
field $\textbf{B}^S$. Dipolar magnetic field is a proper model for
the magnetosphere of a black hole that results from current rings
exterior to the event horizon (Prasanna $\&$ Vishveshwara 1978;
Takahashi $\&$ Koyama 2009)

\[B_r^S=-\frac{3\mu}{4 m^3}\, r^2 \left\{\frac{2m}{r}\left(1+\frac{m}{r}\right)+\ln\left(1-\frac{2 m}{r}\right)\right\} \sin\theta \cos\theta,\]
\[B_{\theta}^S=\frac{3\mu}{4 m^2}\left\{1+\left(1-\frac{2 m}{r}\right)^{-1}+\frac{r}{m}\ln\left(1-\frac{2 m}{r}\right)\right\} \sin^2\theta,\]
wherein, $\mu$ is the dipole  magnetic moment of the central star
that may be expressed in terms of the surface magnetic field $B_s$
and the radius $R$ of the central object as $\mu=B_s R^3$.

To study the magnetic field configuration, the solutions of magnetic
lines of force equations
$$\frac{dr}{B_{(r)}}=\frac{r\, d\theta}{B_{(\theta)}}=\frac{r\sin\theta \, d\varphi}{B_{(\varphi)}},$$
should be analyzed. In order to visualize the field line structure,
it is useful to transform to a Cartesian frame through the usual
relations ($X=r \sin\theta \cos\varphi$, $Y=r \sin\theta
\sin\varphi$, $Z=r \cos\theta$) and then to obtain the corresponding
parametric equations that generate the curves both for the external
magnetic field at infinity
\begin{eqnarray*}
&&X=\frac{r_0}{\sin\theta},\\[.1cm]
&&Z=\frac{r_0}{\sin^2\theta}\cos\theta,
\end{eqnarray*}

\%begin{figure}[htb]
%

and for the disc field
\begin{eqnarray*}
&&X=r_0 \cos (\varphi_0- \beta \, r_0^{-(k-2)} \cot \theta),\\
&&Y=r_0 \sin (\varphi_0- \beta \, r_0^{-(k-2)} \cot \theta),\\
&&Z=r_0 \cot \theta,
\end{eqnarray*}
where $r_0$ and $\varphi_0$ are the constants of integration and
$\beta$ is the ratio of the toroidal magnetic field $B_{(\varphi)}$
to $B_1$. However, in our model $\beta=0$, because of the absence of
$B_{(\varphi)}$. Further, due to the azimuthal symmetry, $\varphi_0$
can be set zero without any loss of generality. Inasmuch as the
meridional structure of the disc is presumed to extend to about
$60^{\circ}$ on either side of the equatorial plane, necessity of
the continuity of the magnetic field lines across the disc boundary
surface (i.e. $\theta=\pi/6$) demands that
\begin{eqnarray}\label{matching condition}
\left(B^D\right)^2|_{r=r_0,\
\theta=\frac{\pi}{6}}=\left(B^S\right)^2|_{r=r_0,\
\theta=\frac{\pi}{6}},
\end{eqnarray}
where
\begin{eqnarray*}
&&\left(B^D\right)^2=B_{(r)}^2+B_{(\theta)}^2,\\[.2cm]
&&\left(B^S\right)^2=\left[B^S_{(r)}\right]^2+\left[B^S_{(\theta)}\right]^2,
\end{eqnarray*}
$r_0$ is the radius where two field lines connect together. Thus,
with the foregoing matching condition (equation \ref{matching
condition}), one can express the constant $B_1$ in terms of the
determined constants as
\[
B_1=\frac{3\sqrt{13}\mu}{8 m^2} \left(\frac{r_0}{2}\right)^{-k} r_0.
\]

Fig. 1 shows a typical profile of the magnetic field structure in
the meridional plane without (Fig. \ref{Magnetic_Line}a) and with
the disc field (Fig. \ref{Magnetic_Line}b). As pointed out by Ghosh
$\&$ Lamb (1978), magnetic lines of force can penetrate the
accretion disc owing to the presence of a finite conductivity.
Because, the value of $\sigma$ is a measure of the rate of slippage
of field lines through the disc plasma. It is seen that, inside the
disc (Fig. \ref{Magnetic_Line}b), the penetrating field lines are
pushed outward as parallel to the $z$ axis. The constant field lines
of the disc are connected with the undistorted external field lines
at the surface of the disc and are continuous at the boundary
surface as postulated earlier. It does not depend on the value of
$k$ as well.

\section{Possible Equilibrium Solution}
To simplify the appearance of the governing equations
(\ref{Continuity equ2}) - (\ref{Azimuthal equ2}), we multiply
equation (\ref{Radial equ2}) by $V^{(r)}$ and equation
(\ref{Meridional equ2}) by $V^{(\theta)}$ and equation
(\ref{Azimuthal equ2}) by $V^{(\varphi)}$ and adding
\begin{eqnarray}\label{Vr*Radial+Vt*Meridional1}
&&\left(\rho+\frac{p}{c^2}\right)\left[\frac{d \left(\frac{V^2}{c^2}\right)}{\left(1-\frac{V^2}{c^2}\right)}+V^{(r)} \frac{\frac{2m}{r^2}}{1-\frac{2m}{r}}\right]+2\, d\left(\frac{p}{c^2}\right)=\nonumber\\
&&-\frac{2}{\sigma u^0
c^2}\left(1-\frac{2m}{r}\right)^{-1}\left[J^{(\varphi)}\right]^2\left\{1-\frac{[V^{(\varphi)}]^2}{c^2}\right\},
\end{eqnarray}
wherein the total velocity $V$ is defined as
\begin{eqnarray*}
V^2=\left[V^{(r)}\right]^2+\left[V^{(\theta)}\right]^2+\left[V^{(\varphi)}\right]^2.
\end{eqnarray*}
Continuity equation (\ref{Continuity equ2}) can be simplified too
\begin{eqnarray}\label{Continuity equ3}
&&\left(\rho+\frac{p}{c^2}\right) {\nabla \cdot \mathbf{\tilde{V}}}+d\left(\rho-\frac{p}{c^2}\right)=\nonumber\\
&&\frac{2}{\sigma u^0 c^2}\left(1-\frac{2m}{r}\right)^{-1}
\left[J^{(\varphi)}\right]^2\left\{1-\frac{[V^{(\varphi)}]^2}{c^2}\right\},
\end{eqnarray}
with a new definition for total velocity as
$\mathbf{\tilde{V}}$$=V^{(r)}\ \hat{r}+\tilde{V}^{(\theta)}\
\hat{\theta}+V^{(\varphi)}\ \hat{\varphi}$, wherein
$\tilde{V}^{(\theta)}=V^{(\theta)}\left(1-\frac{2m}{r}\right)^{-1/2}$.
Right-hand side of these latter two equations (\ref
{Vr*Radial+Vt*Meridional1} and \ref{Continuity equ3}) are similar
with opposite sign. It motivates us to add them and achieve a
simplified equation in terms of the total derivative $d$ with
vanished right-hand side
\begin{eqnarray}\label{Adding equ1}
d
\ln\left[\left(1-\frac{2m}{r}\right)\left(1-\frac{V^2}{c^2}\right)^{-1}\right]
+{\nabla \cdot \mathbf{\tilde{V}}}+d \ln
\left(\rho+\frac{p}{c^2}\right)=0.
\end{eqnarray}

In this equation, all terms have been written in terms of the total
derivative except $\nabla \cdot \mathbf{\tilde{V}}$. If this term
can be written so, then this equation  are integrated simply. To
this aim, we assume $V^{(\theta)}=0$, which is a reasonable
approximation in the case of no outflow production from the disc's
surface (Gu et al. 2009). Accordingly,
$${\nabla \cdot \mathbf{\tilde{V}}}=d\ln \left[r^2 V^{(r)}\right],$$
and equation (\ref{Adding equ1}) is reduced to
\begin{eqnarray}\label{Adding equ2}
d
\ln\left[\left(\rho+\frac{p}{c^2}\right)\left(1-\frac{2m}{r}\right)
\left(1-\frac{V^2}{c^2}\right)^{-1} r^2 V^{(r)}\right]=0,
\end{eqnarray}
that can integrate simply
\begin{eqnarray}\label{C(theta)}
\left(\rho+\frac{p}{c^2}\right)\left(1-\frac{2m}{r}\right)
\left(1-\frac{V^2}{c^2}\right)^{-1} r^2 V^{(r)}=C(\theta),
\end{eqnarray}
where $C(\theta)$ is an arbitrary function of $\theta$. Indeed, it
is the general relativistic definition of the mass accretion rate
$\dot{M}$. Employing the assumption $V^{(\theta)}=0$ in equation
(\ref{Consistency relation}), the radial inflow velocity is acquired
as
\begin{eqnarray}
V^{(r)}=\sqrt{\frac{1-\frac{\left[V^{(\varphi)}\right]^2}{c^2}}{\left(1-\frac{2m}{r}\right)^{-1}+\left(\frac{c}{4\pi\sigma}\frac{B}{r}\right)^2}}\,
\frac{c^2}{4\pi\sigma} \, \frac{B}{r},
\end{eqnarray}
where
$$B=k \left(1-\frac{3m}{r}\right) \left(1-\frac{2m}{r}\right)^{-1}-1+(k-1)\left(1-\frac{3m}{r}\right)^{-1} \cot^2\theta.$$
Contrary to the azimuthal velocity (equation \ref{azimuthal
velocity}), the radial velocity is dependent on conductivity. Here,
a discussion for probable values of the electrical conductivity of
the magnetofluid comes up due to the condition
$|V^{(r)}|<V^{(\varphi)}<c$. We choose the value of conductivity in
the interval
$$2 \times 10^4 \, \frac{1}{s} \leq \sigma \leq 3 \times 10^6 \, \frac{1}{s}.$$
This choice refers to the fact that below the lower bound, the
condition $|V^{(r)}|<V^{(\varphi)}$ is disturbed and above the upper
bound, the radial velocity enters the Newtonian region (i.e.
$|V^{(r)}|<<c$). This range for conductivity is much smaller than
the classical conductivity that is estimated by electron - proton
Coulomb collisions in the Newtonian regime (Kudoh $\&$ Kaburaki
1996)
$$\sigma_{clas} \cong 3\times 10^{12} \left(\frac{T}{10^4 K}\right)^{3/2} \ \frac{1}{s}.$$
Substituting
$$\left(\rho+\frac{p}{c^2}\right)\left(1-\frac{V^2}{c^2}\right)^{-1}=\frac{C(\theta)}{r^2 V^{(r)}} \left(1-\frac{2m}{r}\right)^{-1},$$
in the equations (\ref{Radial equ2}) and (\ref{Meridional equ2}),
the components of the pressure gradient are achieved as
\begin{eqnarray}
\label{dpdr}
&&\frac{\partial p}{\partial r}=-D \, C(\theta)+E \equiv R (r,\theta),\\[.1cm]
\label{dpdtheta} &&\frac{\partial p}{\partial \theta}=F \,
C(\theta)-H \equiv T (r,\theta),
\end{eqnarray}
where
\begin{eqnarray}
\label{D_rel}
&&D=\frac{1}{r^2 \, V^{(r)}}\left(1-\frac{2m}{r}\right)^{-1}\left[V^{(r)}\frac{\partial V^{(r)} }{\partial r}+\frac{m c^2}{r^2}\times\right.\nonumber\\
&&\qquad\left.\left(1-\frac{2m}{r}\right)^{-1}\left\{1-\frac{[V^{(r)}]^2}{c^2}\right\}-\frac{[V^{(\varphi)}]^2}{r}\right],\\[.1cm]
\label{E_rel}
&&E=\frac{1}{c}\left(1-\frac{2m}{r}\right)^{-1/2}B_{(\theta)}J^{(\varphi)}\left\{1-\frac{[V^{(\varphi)}]^2}{c^2}\right\},\\[.1cm]
\label{F_rel}
&&F=\frac{1}{r^2 \, V^{(r)}}\left(1-\frac{2m}{r}\right)^{-1}\cot\theta \, [V^{(\varphi)}]^2,\\[.1cm]
\label{H_rel} &&H=\frac{r}{c}\, B_{(r)}\,
J^{(\varphi)}\left\{1-\frac{[V^{(\varphi)}]^2}{c^2}\right\},
\end{eqnarray}
and $R$ and $T$ are functions of $r$ and $\theta$. A necessary and
sufficient condition for the existence of the solution of the
above-mentioned set of partial differential equations (\ref{dpdr})
and (\ref{dpdtheta}), describing a distribution of pressure in the
fluid, is given by the integrability condition
$$\frac{\partial R}{\partial \theta}=\frac{\partial T}{\partial r},$$
having  the form
\begin{eqnarray}\label{Ctheta}
D \frac{d C(\theta)}{d\theta}+\left[\frac{\partial
D}{\partial\theta}+\frac{\partial F}{\partial r}\right]
C(\theta)=\frac{\partial E}{\partial\theta}+\frac{\partial
H}{\partial r}.
\end{eqnarray}
This condition gives an ordinary differential equation for
$C(\theta)$ which can be solved numerically with an appropriate
boundary condition. Now, with specified $C(\theta)$, one may
determine both the gas pressure as a function of $r$ and $\theta$ by
integrating the equation (\ref{dpdr}) over the radial distance
\begin{eqnarray}
p(r,\theta) = p_0-C(\theta) \int^{r}_{r_0=6 m} D \, d
r+\int^{r}_{r_0=6 m} E \, d r,
\end{eqnarray}
and the gas density through the equation (\ref{C(theta)})
\begin{eqnarray}\label{density_rel}
\rho (r,\theta)=\frac{C(\theta)}{r^2 V^{(r)}}
\left(1-\frac{V^2}{c^2}\right)
\left(1-\frac{2m}{r}\right)^{-1}-\frac{p}{c^2}.
\end{eqnarray}
Here, $p_0$ that has been appeared as an integration constant, is
indeed the pressure of the ISCO ($r_0 = 6\, m$). Since it is
expected that both pressure and density to be positive throughout
the disc, a natural restriction is exerted on $p_0$.


As accretion gives rise to radiation, the equilibrium configuration
is provided only when at the inner layer, the hydrostatic gas
pressure matches the radiation pressure $p_R$, namely $p_0=p_R$.
Under these circumstances, a subcritical regime (i.e.
$\dot{M}<\dot{M}_{cr}=3\times 10^{-8}\frac{M}{M_{\odot}}\,
\frac{M_{\odot}}{year}$) is preferred. Because, at a subcritical
rate of flow of matter into the Roche lobe of a black hole, it may
be assumed that most of the inflowing matter is accreted. However,
at supercritical value of inflow, an effective outflow of matter may
take place under the influence of radiation pressure. It disturbs
the equilibrium configuration of the disc. As a result, to have an
equilibrium structure, we choose the subcritical regime of
accretion. Besides, due to no meridional flow approximation (i.e.
$V^{(\theta)}=0$) employed in our calculations, which is requisite
in the absence of outflows, this regime has been preferred.

At essentially subcritical fluxes $\dot{M}=10^{-12}-10^{-10}
\frac{M_{\odot}}{year}$, the luminosity of the disc is of the order
of $L=10^{34}-10^{36} \, \frac{erg}{s}$. Maximal surface
temperatures are of the order of $T_s=3\times 10^5 - 10^6 K$ in the
inner regions of the disc where most of the energy is released. This
energy is radiated mainly in the ultraviolet and soft X-ray bands,
which are inaccessible to direct observations. When the rate of
accretion increases, the luminosity grows linearly and the effective
temperature of radiation rises.

At fluxes $\dot{M}=10^{-9}-10^{-8} \frac{M_{\odot}}{year}$, disc is
found to be a powerful X-ray source with luminosity
$L=10^{37}-10^{38} \, \frac{erg}{s}$ and an effective temperature of
radiation $T=10^7 - 10^8 K$. It radiates also in the optical and
ultraviolet spectral bands (SS73).

With these observations, it seems suitable to choose
$C(\theta=\frac{\pi}{6})=-10^{-9} \frac{M_{\odot}}{year}$ as a
boundary condition for integrating the differential equation
(\ref{Ctheta}) and $T=10^7 K$ for the temperature of ISCO
($r_0=6m$). Thus, the radiation pressure in that layer is specified
conveniently (i.e. $p_0=\frac{1}{3}\, a\, T^4$, where the radiation
constant $a=7.565767\times 10^{-16} J/m^3 K^4$).


Figs 2 and 3 give the vertical or meridional structure of all
physical variables at some radial distances represented in legend.
As has been deduced from equation (\ref{C(theta)}), mass accretion
rate is just dependent on $\theta$. Accordingly, a proper set of the
free parameters must be chosen so that the mass accretion rate does
not vary significantly with the radial distances (Fig. 2a). Mass
accretion rate is negative as well as radial velocity. Their
negativity indicates the inflow towards the central black hole. Both
radial inflow (Figs 2a and b) and rotation (Fig. 2c) of the disc
slow down from the disc surface ($\theta=\frac{\pi}{6}$) towards the
equator ($\theta=\frac{\pi}{2}$). Pressure shows the similar
behavior too. It falls from the surface towards the equator (Fig.
3a), whereas the density seems to remain nearly constant there (Fig.
3b). This constancy refers to the fact that the variation of density
in radial direction is much larger than that in the meridional
direction (Fig. 4d).


The radial behaviour of the physical variables is plotted in Fig. 4.
Close to the black hole event horizon, the gas temperature and
velocities become extremely high (Popham $\&$ Gammie 1998) and
gradually fall off outwards. Fig. 4 confirms this result as well. It
demonstrates that in radial direction, both radial (Fig. 4a) and
rotational (Fig. 4b) velocities of the disc become faster inwards.
Pressure and density are the descending functions of the radial
distance as well (Figs 4c and d). The density falls off rapidly as
$r$ increases whereas pressure tends to remain constant after an
initial decrease.

The effect of free parameters $\sigma$ and $k$ on some impressible
physical variables is investigated by plotting them as a function of
$r$ for different values of $\sigma$ (Fig. 5) and $k$ (Fig. 6).
Rotational velocity is not affected by these parameters. However,
radial inflow velocity slows down with rising them (Figs 5a and 6a).
For lower values of $\sigma$, the pressure decrease outward is
minimal. As $\sigma$ becomes larger, this difference becomes
appreciable and also the pressure drops off (Fig. 5b) contrary to
the behavior of the density. The disc becomes denser with rising the
conductivity (Fig. 5c) and free parameter $k$ (Fig. 6b).

%
%

Fig. 7 shows the meridional dependency of the pressure and density
with respect to the angular momentum parameter $l$. As disc rotates
faster, both pressure (Fig. 7a) and density (Fig. 7b) increase.
However, ascending the pressure with $l$, is just significant on the
surface layers and gradually this sensitivity diminishes towards the
equator. Furthermore, higher the value of $l$, larger the variation
of density with $\theta$.

Isodensity contours for different values of $\sigma$ and $k$ are
plotted in Figs 8 and 9 respectively. For lower values of these
parameters, the related isodensity contours occur in nearer radial
distances to central star.

%

Meridional flow pattern for different values of $\sigma$ and $k$ is
depicted in Figs 10 and 11 respectively. The flow is represented by
the arrows at a grid of points in the X-Z plane indicating the
direction of streamlines. As mentioned before, the conductivity
ascent leads to decelerate the radial inflow velocity, so that the
rotation would be much faster than the radial inflow for large
values of $\sigma$. We would also have a rotating non-accreting
equilibrium configuration around the central star (Fig. 10).
Opposing to this behaviour is observable in Fig. 11 with lowering
$k$. Because in this case, radial inflow velocity becomes faster
than rotational velocity in such a way that for $k=-3$, inflow
occurs as pure free fall without any rotation.

%

\section{Newtonian Limit and Comparison with the TPD90's Solutions}

To convince about the correctness of our solutions and also
comparing with those of TPD90, one may employ the Newtonian limit
(i.e. $m<<1$ and $\frac{V}{c}<<1$) on the equations (\ref{azimuthal
velocity}) - (\ref{Consistency relation})
and acquires their Newtonian counterparts as follows:\\
Azimuthal velocity
\begin{eqnarray}\label{azimuthal velocity_N}
V^{(\varphi)}=\frac{\tilde{L}}{r\sin\theta},
\end{eqnarray}
Components of electromagnetic field
\begin{eqnarray}
B_{(r)}=-B_1 \, r^{k-1} \sin^{k-1}\theta \, \cos\theta,
\end{eqnarray}
\begin{eqnarray}
B_{(\theta)}=B_1 \, r^{k-1} \sin^{k}\theta,
\end{eqnarray}
\begin{eqnarray}
E_{(r)}=\frac{\tilde{L}}{c} B_1 \, r^{k-2} \sin^{k-1}\theta,
\end{eqnarray}
\begin{eqnarray}
E_{(\theta)}=\frac{\tilde{L}}{c} B_1 \, r^{k-2} \sin^{k-2}\theta \,
\cos\theta.
\end{eqnarray}
and current density
\begin{eqnarray}
J^{(\varphi)}=(1-k) \frac{c}{4\pi} B_1 \, \left(r \sin
\theta\right)^{k-2},
\end{eqnarray}
\begin{eqnarray}
J^{(t)}=(1-k) \frac{\tilde{L}}{4\pi} B_1 \, \left(r \sin
\theta\right)^{k-3},
\end{eqnarray}
Consistency equation
\begin{eqnarray}
V^{(r)}+\cot\theta \, V^{(\theta)}=(k-1) \frac{c^2}{4\pi\sigma}
\frac{1}{r \sin^2\theta}.
\end{eqnarray}
According to TPD90's notation, $\tilde{L}^2=l\, M\, G\, r_{in}$,
wherein the inner radius locates in $r_{in}=15 \, m$. Tilde symbol
is written over the Newtonian counterparts of the variables to
distinguish them from the relativistic ones. Considering the
condition $V^{(\theta)}=0$, the radial inflow velocity is achieved
too
\begin{eqnarray}
V^{(r)}=(k-1) \frac{c^2}{4\pi\sigma} \frac{1}{r \sin^2\theta}.
\end{eqnarray}
As we know, $V^{(r)}$ must be negative to indicate inflows. It
exerts an upper bound on the integer $k$ as $k\leq0$. In addition,
the condition $|V^{(r)}|<<c$ leads to
$$\sigma>>(1-k) \frac{c}{4\pi}\frac{1}{r\sin^2\theta}.$$
Newtonian counterpart of the equation (\ref{density_rel}) provides a
relation for density as
\begin{eqnarray*}
\rho=\frac{\tilde{C}(\theta)}{r^2
|V^{(r)}|}=\frac{\tilde{C}(\theta)}{1-k} \, \frac{4\pi \sigma}{c^2}
\, \frac{\sin^2\theta}{r}.
\end{eqnarray*}
Although TPD90 assumes that the mass accretion rate
$\tilde{C}(\theta)$ is constant and denotes it by $\dot{M}$
\begin{eqnarray}\label{density_N}
\rho=\frac{\dot{M}}{1-k} \, \frac{4\pi \sigma}{c^2} \,
\frac{\sin^2\theta}{r}.
\end{eqnarray}
Equations (\ref{D_rel}) - (\ref{H_rel}) in Newtonian limit are
simplified as
\begin{eqnarray*}
&&\tilde{D}=\frac{1}{r^2 |V^{(r)}|}\left[V^{(r)} \frac{\partial V^{(r)}}{\partial r}+\frac{M G}{r^2} -\frac{\left[V^{(\varphi)}\right]^2}{r}\right]\nonumber\\[.1cm]
&&\quad=\frac{c^2}{4\pi \sigma} \frac{(k-1)}{r^4 \sin^2\theta}+\frac{4\pi \sigma}{c^2}\frac{1}{(1-k)}\left[\frac{M G}{r^3} \sin^2\theta-\frac{L^2}{r^4}\right],\qquad\\[.2cm]
&&\tilde{E}=\frac{1}{c}\, B_{(\theta)} J^{(\varphi)}=\frac{B_1^2}{4\pi}\ (1-k) \, r^{2k-3} \sin^{2k-2}\theta\nonumber\\[.1cm]
&&\qquad\qquad\qquad\ \ \ =-\frac{\partial}{\partial r}\left(\frac{B_p^2}{8\pi}\right),\\[.2cm]
&&\tilde{F}=\frac{\cot\theta}{r^2 V^{(r)}} \left[V^{(\varphi)}\right]^2=\frac{4\pi\sigma}{c^2}\frac{L^2}{k-1}\frac{\cot\theta}{r^3},\\[.2cm]
&&\tilde{H}=\frac{r}{c}\, B_{(r)} J^{(\varphi)}=-\frac{B_1^2}{4\pi}\ (1-k) \, r^{2k-2} \sin^{2k-3}\theta \cos\theta\nonumber\\[.1cm]
&&\qquad\qquad\qquad\quad=\frac{\partial}{\partial
\theta}\left(\frac{B_p^2}{8\pi}\right),
\end{eqnarray*}
where poloidal magnetic field $B_p$ is defined as
$$B_p^2=B_{(r)}^2+B_{(\theta)}^2=B_1^2 (r \sin\theta)^{2k-2}.$$
Under these circumstances, by introducing the effective pressure
$$\bar{p}=p+\frac{B_p^2}{8\pi},$$
equations (\ref{dpdr}) and (\ref{dpdtheta}) become more concise
\begin{eqnarray}\label{dpdr2}
&&\frac{\partial \bar{p}}{\partial r}=-\tilde{D} \, \dot{M} \equiv \tilde{R}(r,\theta),\\[.1cm]
&&\frac{\partial \bar{p}}{\partial \theta}=\tilde{F} \, \dot{M}
\equiv \tilde{T}(r,\theta).
\end{eqnarray}
Integrating the equation (\ref{dpdr2}) over the radial distance
yields
\begin{eqnarray*}
\bar{p} \, (r,\theta) = -\dot{M} \int^{r}_{r_{in}=15 m} \tilde{D} \,
d r,
\end{eqnarray*}
then, the gas pressure can be obtained
\begin{eqnarray}\label{p2}
p=p_0&+&\left\{\frac{2\pi\sigma G M \dot{M}}{(1-k) c^2} \frac{\sin^2\theta}{r^2}-\frac{1}{3 r^3}\left[\frac{(1-k) \dot{M} c^2}{4\pi \sigma \sin^2\theta}+\frac{4\pi \sigma \dot{M} L^2}{(1-k) c^2}\right]\right\}\nonumber\\[.1cm]
&-&\frac{B_p^2}{8\pi}.
\end{eqnarray}
These solutions (equations \ref{azimuthal velocity_N} -
\ref{density_N} and \ref{p2}) are in exact agreement with those
obtained by TPD90. Now, it deserves to compare the relativistic and
Newtonian solutions both qualitatively and quantitatively. In close
vicinity of a black hole, the Newtonian description is only a rough
approximation and it is natural to expect the effects of space–time
curvature there. Thus, employing the Newtonian approximation is just
permissible on farther radial distances relative to central compact
object. That is why the ISCO is placed farther for Newtonian regime.
Another quantitative difference rests on the range of velocities
(Figs 12a,b and 13a,b). It demands two different ranges for the free
parameters $\sigma$ and $l$ for both regimes (Table 1). In Newtonian
regime, this allowed range of free parameters results in a huge
reduction in the density of the disc relative to the relativistic
regime (Fig. 12c). Besides that the density has an obvious
quantitative difference in both regimes, it behaves qualitatively
different as well. The descending radial gradient density in
relativistic regime is steeper than Newtonian regime (Fig. 13c). The
pressure acts qualitatively different in both regimes as well as
density (Figs 12d and 13d). In relativistic regime, pressure
diminishes from the layer surface toward the equator in meridional
direction. However, exactly contrary to this behaviour can be seen
in Newtonian regime (Fig. 12d). Moreover, in relativistic regime, in
radial direction, pressure decreases rapidly in the inner radial
distances, then steadies itself and reaches an almost constant value
as one goes outwards in the disc. Whereas in Newtonian regime,
pressure falls off rapidly (Fig. 13d). Figs 14 and 15 have been
depicted to illustrate the role of $\sigma$ and $k$ in Newtonian
regime respectively. With respect to these free parameters, all
impressible physical variables behave similarly in both regimes
(Figs 14a,c and 15), except the pressure. In Newtonian regime, quite
the opposite to the relativistic regime, pressure rises with an
increase in conductivity (Fig. 14b).
\begin{table*}\label{Table 1}
 \centering
 \begin{minipage}{140mm}
 \caption{Range of free parameters in Newtonian and Relativistic Regimes.}
 \begin{tabular}{@{}cc@{}}
 \hline
   Newtonian Regime & Relativistic Regime\\
 \hline
$|\frac{V}{c}|<<1$       &       $|\frac{V}{c}|<1$ \\[.3cm]
$\sigma >> (1-k) \frac{c}{4\pi}\frac{1}{r\sin^2\theta}\approx 5
\times 10^4 \frac{1}{s}\qquad\qquad$ &
$2 \times 10^4 \, \frac{1}{s} \leq \sigma \leq 3 \times 10^6 \, \frac{1}{s}$  \\[.3cm]
$l=0.1$ & $0.1\leq l \leq 1.5$  \\[.2cm]
$r_{in}=15 \, m$ & $r_{ISCO}=6 \, m$\\
\hline
\end{tabular}
\end{minipage}
\end{table*}

\input{epsf}
\section{Conclusion}
In this paper, we find a self-consistent solution of fully
relativistic equations for a thick disc around a compact object
having the radial and azimuthal components of the flow velocity
non-zero. In general, it is not easy to solve this set of coupled
partial differential equations without invoking the simplifying
assumptions. Mainly because these equations are so complicated and
highly non-linear, especially when the disc is magnetized. We have
included a finite conductivity for the plasma and have ignored the
shear viscosity and self-gravity of the disc. Despite the great
simplifications coming from these assumptions, the scenario is still
physically reasonable and non-trivial. Few variety of degrees of
freedom can be captured and the free parameters can be conveniently
chosen in order to describe an astrophysically relevent situation.

Using Ohm law explicitly, we have derived a self-consistent
equilibrium solution for a plasma disc in the presence of an
external stellar dipolar magnetic field along with a
self-consistently generated poloidal magnetic field of the disc.
This class of solutions is somewhat a general relativistic
generalization of the Newtonian solutions obtained by TPD90. The
space-time curvature produced by the strong gravitational field of
the central body modifies the magnetic fields and other physical
variables relative to the Newtonian case. However, similar to
Newtonian model of TPD90, the magnetic field lines inside the disc
are constant in nature.

The meridional structure of our magnetized thick disc is mainly
determined by the force balance of the vertical component of the
plasma pressure gradient, magnetic and centrifugal forces rather
than that of gravity and gas pressure like in the standard viscous
disc model. The existence of such structure, in fact, encourages one
to look for generalizations of the analysis to cases having
meridional flow (i.e. $V^{\theta}\neq0$). Of course, such more
complicated choice requires more complicated calculations. This
might be suggestive of generating the jets from the disc. Such
analyses of the thick accretion disc dynamics are in progress, but
are beyond the scope of the present paper.

Even if such resistive tori with foregoing particular
characteristics do not exist in nature, the exact solutions
presented here can still be useful for numerical general
relativistic MHD, which has attracted a lot of interest (Koide,
Shibata $\&$ Kudoh 1999; Komissarov 2004; Anton et al. 2006).

\section*{Acknowledgments}

The author would like to thank the anonymous referee for very useful
comments and valuable points to improve the presentation of the
paper. This work has been carried out under the financial support of
Islamic Azad University, Fars Science and Research Branch and Shiraz
Branch as a research proposal.

\end{document}